\documentclass[noshowpacs,twocolumn, amsmath, amssymb, aps, prx,superscriptaddress]{revtex4-2}

\usepackage{xcolor}

\usepackage[colorlinks=true]{hyperref}

\usepackage{svg}
\usepackage[T1]{fontenc}

\usepackage{bibunits}
\defaultbibliographystyle{naturemag}
\defaultbibliography{Netrunners}

\usepackage{orcidlink}

\newcommand{\cnn}{Centre de Nanosciences et de Nanotechnologies, Universit\'e Paris-Saclay,  CNRS, Palaiseau, France}

\newcommand{\quandela}{Quandela SAS, Massy, France}

\newcommand{\maxplanck}{Max Planck Institute for the Science of Light, Erlangen, Germany}

\newcommand{\massachusetts}{College of Information and Computer Sciences, University of Massachusetts Amherst, Amherst, Massachusetts, USA}

\begin{document}

\title{Benchmarking quantum key distribution by mixing single photons and laser light}

\author{Yann Portella\,\orcidlink{0009-0006-6362-9246}}
\author{Petr Steindl\,\orcidlink{0000-0001-9059-9202}}
\author{Juan Rafael Álvarez\,\orcidlink{0000-0003-3411-8237}}
\affiliation{\cnn}

\author{Tim Hebenstreit}
\affiliation{\maxplanck}

\author{Aristide Lemaître\,\orcidlink{0000-0003-1892-9726}}
\author{Martina Morassi\,\orcidlink{0000-0001-6472-2916}}
\affiliation{\cnn}

\author{Niccolo Somaschi}
\affiliation{\quandela}

\author{Loïc Lanco\,\orcidlink{0000-0003-3607-7539}}
\affiliation{\cnn}

\author{Filip Rozp\k{e}dek\,\orcidlink{0000-0002-2755-4623}}
\affiliation{\massachusetts}

\author{Pascale Senellart\,\orcidlink{0000-0002-8727-1086}}
\affiliation{\cnn}

\author{Dario A. Fioretto\,\orcidlink{0000-0003-3829-4000}}
\affiliation{\cnn}
\affiliation{\quandela}

\date{\today}

\begin{abstract}

Quantum key distribution is a key application of quantum mechanics, shaping the future of privacy and secure communications. Many protocols require single photons, often approximated by strongly attenuated laser pulses. Here, we harness the emission of a quantum dot embedded in a micropillar and explore a hybrid approach where the information is encoded on a mixture of single photons and laser pulses. We derive a phenomenological analysis of the configuration where both sources of light are mixed incoherently to perform the BB84 protocol, showing nearly perfect matching between theory and experiment. This provides a flexible technology compensating limited collected brightnesses of single-photon sources as well as a thorough investigation of single-photon statistics advantage scenarios over Poisson-distributed statistics. Explicitly, our model highlights an efficiency threshold for unconditional advantage of single photons over laser along with insights on the interplay between single-photon purity and collected brightness in the performances of BB84.

\end{abstract}

\maketitle

\section*{Introduction}

\begin{bibunit}

\begin{figure*}
    \centering
    \includegraphics[scale = 0.56]{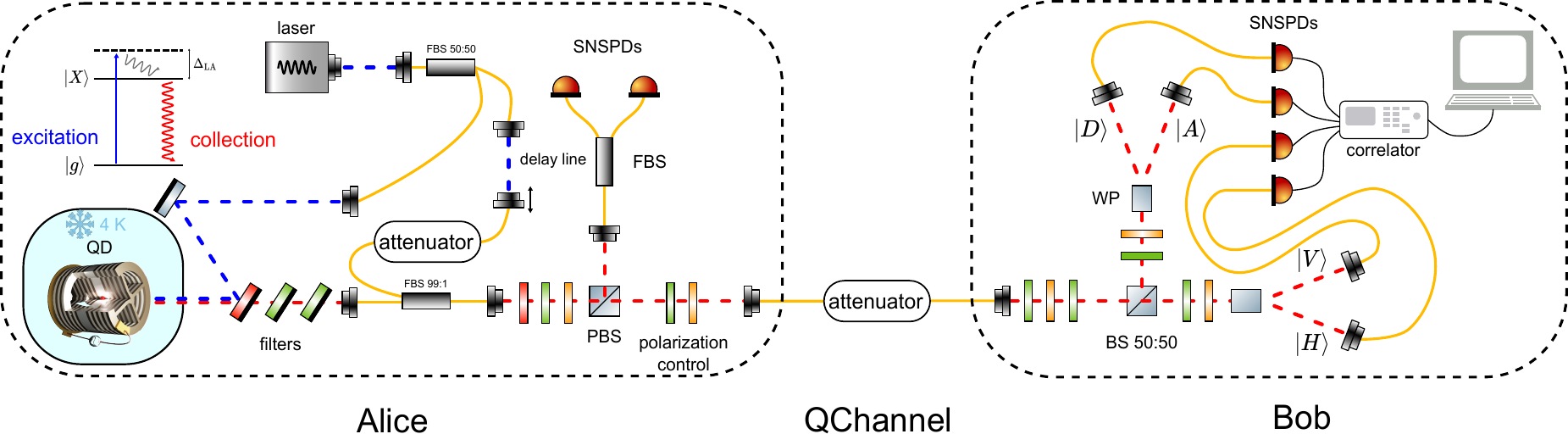}
    \caption{\textbf{BB84 implementation with a QDS.} Photons are produced by exciting the QD, operating in a 4 K cryostat, with blue-detuned laser light by $\Delta_{\text{LA}}$. The excitation laser is split into two different paths using a 50:50 fibered beam-splitter (FBS). One path is used to excite the QD, and the laser light is later suppressed with three bandpass filters with two (green) mounted on motorized stages. The second path contains a delay line to temporally match laser pulses and single photons that are recombined with a 99:1 fibered coupler. The photons are linearly polarized using a linear polarizer (red) and a quarter (green) and a half waveplate (yellow). A polarizing beam splitter (PBS) allows a small pick-off of the photons into a Hanbury Brown and Twiss (HBT) measurement setup. The four BB84 states are encoded using another pair of half and quarter waveplates. After the variable optical attenuator (VOA) used to emulate losses over long fiber links, a polarization analyser setup enables randomized measurements in two orthogonal bases: $\{|H\rangle, |V\rangle\}$ and $\{|D\rangle, |A\rangle\}$ with two waveplates and one Wollaston prism (WP) on each arm. The events measured on superconducting nanowire single-photon detectors (SNSPDs) are monitored and processed using a time tagging device.}
    \label{fig:setup}
\end{figure*}

Quantum communications protocols and quantum networks tackle many concerns about the future of privacy and are a feasible approach to scale quantum computing by interconnecting remote quantum processors \cite{wehner_quantum_2018}. Quantum key distribution (QKD) is a primitive of quantum communication where two legitimate users trusting each other (Alice and Bob) distribute a key, with the secrecy of this exchange being guaranteed by the laws of physics \cite{xu_secure_2020}.

To ensure security, the implementation of QKD protocols require qubits to be carried via single photons, benefiting from the no-cloning theorem. However, producing high-performance single-photon sources (SPS) remains technically challenging. As a result, significant theoretical and experimental progress has been made using Poisson-distributed sources (PDS) such as attenuated laser pulses, as practical substitutes for ideal single photons \cite{liao_satellite--ground_2017, liao_satellite-relayed_2018, grunenfelder_simple_2018}. The drawback of PDS is the occurrence of multiphoton emissions, which introduce security loopholes that an eavesdropper (Eve) can exploit -- for example, with a photon-number splitting (PNS) attack \cite{lutkenhaus_quantum_2002}. The fundamental vulnerability induced by the multiphoton components can be suppressed by reducing the mean photon number, which inevitably lowers the amount of transmitted photons from Alice to Bob. This significant performance reduction can be overcome with techniques such as using decoy states \cite{lo_decoy_2005} and by monitoring photon number correlations to reduce uncertainties on the source's parameters \cite{dynes_testing_2018}. With near ideal single-photon purity, SPS do not suffer from this problem and could allow longer communication distance \cite{waks_security_2002}. Still, the full advantage of SPS requires high photon collected brightness, a requirement that has long remained technologically challenging.

Recently, semiconductor quantum-dot sources (QDS) have demonstrated high performances, generating single photons on demand. The collected brightness has been constantly increased, reaching record value in terms of fiber brightness (probability to get a photon on demand) up to 50-70\,\% \cite{somaschi_near-optimal_2016, tomm_bright_2021, ding_high-efficiency_2025}. This comes with a near unity single-photon purity above 97\,\%. These features have recently allowed reaching high key rates and cover long distances \cite{bozzio_enhancing_2022, zahidy_quantum_2024,  yang_high-rate_2024}. It has also been proven experimentally that QDS can surpass the fundamental limit achievable with PDS in free space demonstrations \cite{zhang_experimental_2025}. Lastly, QDS are also now integrated in plug-and-play commercial devices \cite{margaria_efficient_2025}. These milestones represent a fundamental landmark for the technology, opening the way to practical quantum communication studies using QDS. In the present work, we explore a route to reach higher performances in QKD by taking advantage of both QDS and PDS. We implement a proof-of-concept of the polarization-encoded BB84 protocol \cite{bennett_quantum_2014} with the additional ability to mix photon-number statistics of single photons and laser light at the transmitter output. We show that such incoherent mixing of those sources can be leveraged to increase the secret key rate (SKR) by tuning the ratio of the mixing, depending on the QDS brightness and photon losses. This shows a new method to increase secret key rates in realistic scenarios, and paves the way towards hybrid approaches for quantum communications as well as better understanding of advantage conditions for QDS.

\section*{Results}

\subsection*{Experimental protocol}

We investigate a BB84-like setup \cite{bennett_quantum_2014} where Alice and Bob distribute a secret key composed of bits of information encoded in the polarization of photons in two bases: \{$|H\rangle$,$|V\rangle$\} and \{$|D\rangle$,$|A\rangle$\}. The bases are later reconciled publicly: whenever the preparation and measurement bases match, a raw random bit is shared between the two users.

We implement a proof-of-concept setup for BB84 using three modules shown in Fig. \ref{fig:setup}. A first module (Alice) contains the generation of single photons, their spectral selection, incoherent mixing with the excitation laser, and subsequent polarization control. The semiconductor quantum dot (QD) used in this demonstration is embedded in a micropillar cavity where the Purcell effect allows for efficient photon collection \cite{somaschi_near-optimal_2016, thomas_bright_2021}. The source is placed in a cryostat at approximately 4K. A neutral exciton from which we consider only one of the linearly polarized optical transition (energy level diagram in Fig. \ref{fig:setup}), is excited using 10 ps laser pulses, blue-detuned by $\Delta_{\text{LA}} = $ 0.8 nm with respect to the QD transition, at a repetition rate of 81.96 MHz. This excitation scheme enables an efficient population transfer to the excited state assisted by the interaction with acoustic phonon during the excitation process \cite{thomas_bright_2021}. The excitation laser is precisely aligned in polarization to one of the two dipoles of the neutral exciton, guaranteeing polarized monochromatic single-photon emission. The frequency difference between the laser pulse and the single photons offers the practical advantage of easily separating the two via commercially available spectral filters. Furthermore, the bandwidth of these filters can be tuned by changing their tilt angle with respect to the incident beam. We use optimized filtering (see Supplementary Material) to maintain high brightness and single-photon purity with a detected count rate of 8 MHz and a second-order intensity correlation $g^{(2)}(0) = 1.2 \%$ measured in-fiber, directly at the output of the filtering stage. Additionally, our excitation scheme naturally suppresses any residual coherence between photon-number components \cite{thomas_bright_2021} leading to a wider stability over excitation power fluctuations. This allows for a stable, long-term characterization of the source parameters \cite{maring_versatile_2024}, essential for guaranteeing security.

The collected single photons are recombined with the laser pulses that were previously separated from the excitation path with a 99:1 fibered coupler. The laser pulses coming from this separate path are synchronized with the single photons using a delay line. Their mean photon number $\mu_{\text{laser}}$ is controlled with a variable optical attenuator. The polarization of the photons coming from the two inputs is aligned with a polarizer and controlled with a set of half and quarter waveplates to reflect a small fraction of the photons with a polarized beamsplitter. These reflected photons are sent towards a Hanbury Brown and Twiss (HBT) experiment where the $g^{(2)}(0)$ of the mixed light stream is measured. Finally, the polarization of the transmitted photons, used to encode information, is set using an additional set of waveplates at the output of the transmitter.

The photons are then sent into the quantum channel (QChannel) whose losses over long-fiber links are emulated locally using a voltage-controlled attenuator. Changing the attenuation of the quantum channel allows measuring the key figures of merit for QKD corresponding to several distances between Alice and Bob.

Ultimately, the photons are measured at a polarization analysing setup (Bob) in two orthogonal bases. A first set of waveplates is used to compensate for polarization drifts happening along the quantum channel. A balanced beam-splitter randomly directs each photon into one of the measurement basis: \{$|H\rangle$, $|V\rangle$\} or \{$|D\rangle$, $|A\rangle$\}. Each output path includes two waveplates and a Wollaston prism, which together enable a polarization analysis in the given basis. All four possible polarization states are thus collected in four different paths and finally measured using four superconducting nanowire single-photons detectors (SNSPDs). This setup enables the measurement of all relevant parameters for assessing the SKR -- namely the raw count rates measured by Bob, the quantum bit error rate (QBER), and the single-photon purity of the transmitted light ($g^{(2)}(0)$).

\subsection*{Security bounds with imperfect single-photon sources}

In our implementation, errors occur when an event is measured on the detector corresponding to the state orthogonal to Alice's encoding. In practice, this is determined by our capability to correctly distinguish two orthogonal polarization states in two different bases. In our experiment, we checked that the same amount of errors could be observed by sending $|V\rangle$ instead of $|H\rangle$ or by encoding and measuring in the $\{|D\rangle, |A\rangle\}$ basis. Hence, we only encode for the $|H\rangle$ state and consider the errors only in the \{$|H\rangle$, $|V\rangle$\} measurement basis, based upon the assumption that error rates will be the same in the other detection basis. The total error rate $e$ in the $\{|H\rangle, |V\rangle\}$ basis, including dark counts, depends highly on the probability $e_{\text{d}}$ for each photon measured in the correct basis to have entered the wrong detection path. From the data measured using single photons, we inferred $e_{\text{d}} = 0.8 \%$. The case of higher $e_{\text{d}}$ is explored in the Supplementary Material. Beyond this, the parameters affecting the amount of secret key shared at the end of the protocol are the number of multi-photon emissions, and the number of events detected by Bob. With its high single-photon purity, determined by the value of $g^{(2)}_{\text{QD}}(0)$, and relatively small error rate due to the local nature of this demonstration, our QDS is used to generate non-zero secret key for attenuations of up to 30 dB. However, the fiber collected brightness at the output of Alice $\mathcal{B}$ of 4.09\,\% limits the number of generated secret bits. All relevant parameters of the experiment are summarized in Table \ref{tab:parameters}.

\begin{table}[h]
\begin{tabular}{|c| c|} 
 \hline
 $e_{\text{d}}$ & 0.8\,\% \\
 \hline
 QDS single-photon purity 1 - $g^{(2)}_{\text{QD}}(0)$ & 98.8\,\% \\ 
 \hline
 Repetition rate & 81.96 MHz \\
 \hline
 Collected brightness at the output of Alice $\mathcal{B}$ & 4.09\,\% \\
 \hline
 Error-correcting factor $f_{\text{EC}}$ & 1.2 \\
 \hline
 Dark-count rate & 196 Hz \\
 \hline
\end{tabular}
\caption{Parameters of the experiment}
\label{tab:parameters}
\end{table}

Formally, as in the framework of other experimental works \cite{morrison_single-emitter_2023, zahidy_quantum_2024, kupko_tools_2020}, a key metric of a QKD system's performances is the estimation of an asymptotic lower bound on the generated key. We place ourselves in the worst case scenario, described by the GLLP security proof, where every multi-photon component is leaked to Eve as a tagged event \cite{gottesman_security_2004}. This leads to a lower bound on the asymptotical SKR for BB84:

\begin{equation}
    \text{SKR} \geqslant \frac{p_{\text{click}}}{2} \left( A \left(1 - H_2\left(\frac{e}{A} \right) \right) - f_{\text{EC}} \left( e \right) H_2\left(e \right) \right)\,,
    \label{eq:SKR}
\end{equation}
where $H_2(x)$ is the binary entropy function:
\begin{equation}
   H_2(x) = -x\log_2(x) - (1 - x)\log_2(1-x)\,.
   \label{eq:H2}
\end{equation}

In Eq. \ref{eq:SKR}, $p_\mathrm{click}$ is the detected photon rate at the end node and the factor $\frac{1}{2}$ takes into account the sifting, the event ratio where Alice prepares and Bob measures in the same basis. The error rate $e$ measures the error within the same measurement basis. The term $f_\mathrm{EC} \left( e \right) H_2\left(e \right)$ accounts for the error correction; the error correcting function $f_\mathrm{EC}(e) = 1.2$ is considered constant in the range of error rates we observe according to \cite{lutkenhaus_security_2000}. Finally, the other term $A \left(1 - H_2\left(\frac{e}{A} \right) \right)$ accounts for a further compression of the key due to privacy amplification with $A$ being the single-photon component of the beam at the Alice's output. The value of $A$ plays an important role as more stringent privacy amplification is needed as the multi-photon component increases. We can estimate a lower bound for $A$ in an agnostic way from the photon-number distribution using the second-order correlation function $g^{(2)}(0)$ and the mean-photon number $\mu$ with:

\begin{equation}
    A = \frac{p_{\text{click}} - p_{\text{m}}}{p_{\text{click}}} \quad\text{and}\quad p_{\text{m}} \leqslant g^{(2)}(0) \frac{\mu^2}{2}\,,
    \label{eq:pm}
\end{equation}

where $p_\text{m}$ is the probability to generate a multi-photon state. Eq. \ref{eq:SKR} shows explicitly the parameters impacting the SKR. The estimation in Eq. \ref{eq:SKR} could be further refined introducing more advanced techniques such as decoy states \cite{hwang_quantum_2003, lo_decoy_2005}, and analysis methods \cite{dynes_testing_2018, kamin_improved_2024}. To estimate $A$ and deduce the SKR from our experimental data, we infer the mean photon numbers due to single photons and laser at the output of Alice by measuring dark counts (events not triggered by any BB84 qubit detection), total events and events triggered by QDS emissions (blocking the laser path in Fig. \ref{fig:setup}). Although the estimation for $p_\text{m}$ shown above is quite accurate for SPSs, as they have very low probability of generating states with more than 3 photons, it needs to be refined in the case where the amount of laser we are sending in is not negligible. Details on the derivation of said mean photon numbers and the theoretical model of hybrid statistics can be found in the Supplementary Material.

\begin{figure}
    \centering
    \includegraphics[scale = 0.55]{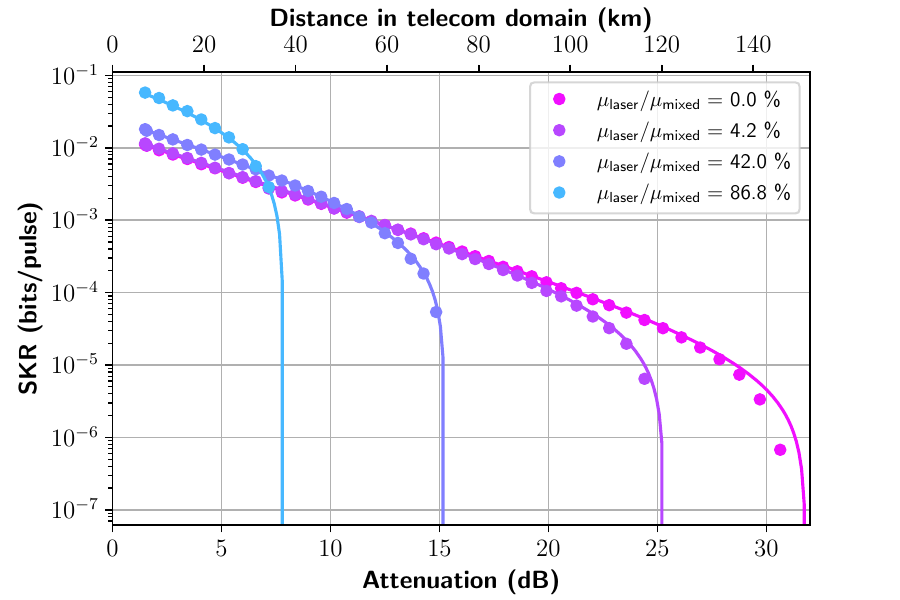}
    \caption{\textbf{Distance scalings of the SKR for different mixed statistics.} The secret key rate in bits per pulse is measured over attenuation for different mixed statistics, corresponding to different amounts of poissonian light sent among single photons to Bob (experiment : data points, simulation : lines). The ratio of the photons coming from the two types of statistics is given by the ratio of their mean photon numbers. The attenuation is mapped to distance assuming direct operation in the telecom C-band, using commercial fibers with linear attenuation $\alpha = $ 0.21 dB/km.}
    \label{fig:scalings}
\end{figure}

In Fig. \ref{fig:scalings}, we plot the obtained SKR as we progressively mix single photons with different intensities of laser light as a function of increasingly larger attenuation. To quantify the amount of laser in this incoherent mixture we take the ratio of the mean photon number coming from the laser $\mu_\text{laser}$ over the total mean photon number $\mu_\text{mixed}$. We find that compared to laser pulses, our QDS is limited in the SKR at short distances because of the low collected brightness as discussed in the following. Fig. \ref{fig:scalings} evidences that this limitation can be overcome by mixing QDS single photons with Poissonian light of low intensity and by that increasing the SKR at low distances.

Even though adding laser reduces the maximal distance for which Alice and Bob can share a key (i.e. the secret key rate is positive), it can be also noticed that the secret key rate increases by up to one order of magnitude at short distances compared to using single photons alone. Conversely, for larger distances, increasing the laser fraction results in increased multi-photon states, compromising the security. This effect is evidenced in Eq. \ref{eq:SKR} within the privacy amplification term where more counts coming from multi-photon emissions leads to more distillation needed to get a secure key.

\begin{figure*}
    \centering
    \includegraphics[scale = 0.50]{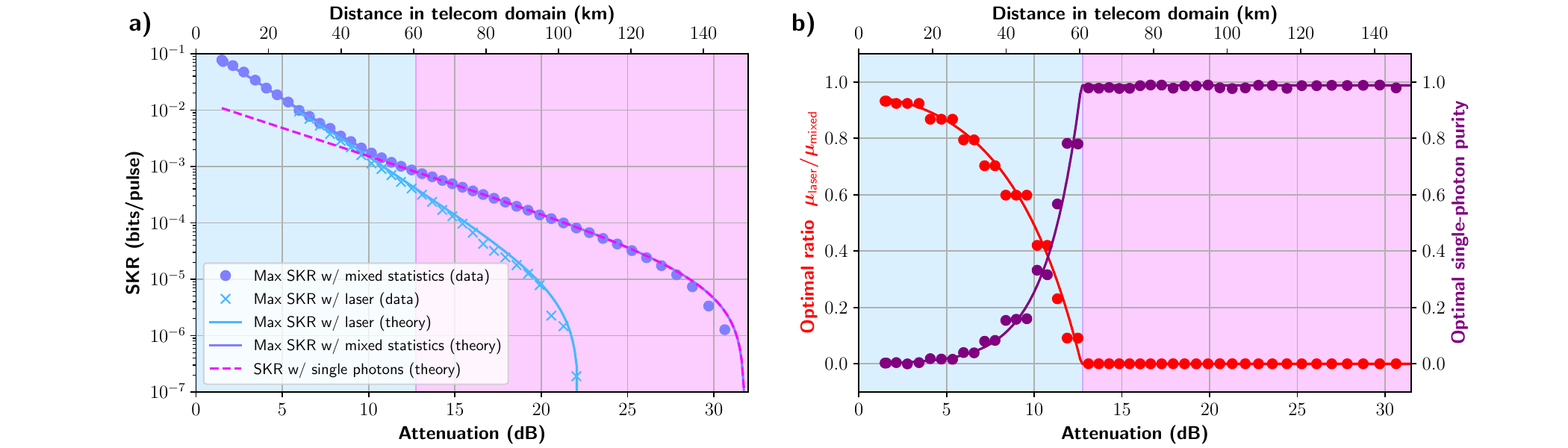}
    \caption{\textbf{SKR distance scalings and optimized parameters of the mixed statistics as a function of attenuation. a)} The SKR is measured over several attenuations for multiple amounts of Poissonian light used by Alice. The maximum SKR obtained for each attenuation is plotted (purple dots) and compared to the theoretical model (purple solid line). As comparison, the SKR distance scaling using only single photons from our source is plotted in pink, and the measured SKR using only laser is plotted with the blue crosses, compared to the theoretical curve in blue solid line. \textbf{b)} Comparison of the experimental (symbols) optimal relative laser mean photon number (red) and single photon purity $\mathcal{P} = 1 - g^{(2)}(0)$ (purple) with our model (solid lines).}
    \label{fig:optimizedscaling}
\end{figure*}

In our experiment, we span mean photon number ratios from 0 \% (no laser) to 86.8 \% (Fig. \ref{fig:scalings}), showing the ability to select from a wide range of mixed statistics. This ability to control both the attenuation of the quantum channel and of the mixed laser pulses allows to scan for the maximum SKR that can be obtained for a given distance between Alice and Bob over the range of mean photon numbers from the laser. The data is shown in Fig. \ref{fig:optimizedscaling} where we look at the advantage provided by the hybrid statistics from two different perspectives. In Fig. \ref{fig:optimizedscaling}a., we compare the SKR maximized by mixing \textit{for each distance} different but optimal $\mu_\text{laser}$ with the expected SKR obtained with our QDS and with laser. The optimal photon statistics evolution of the hybrid statistics parameters maximizing the SKR are shown in Fig. \ref{fig:optimizedscaling}b.

From these observations, we identify two regimes in the distance scaling of the secret key rate. Above 12 dB (colored in light violet background in Fig. \ref{fig:optimizedscaling}a.), the optimal distance scaling of SKR is reached using only single photons from the QDS. Interestingly, at shorter distances, below 12 dB (colored in light blue background in Fig. \ref{fig:optimizedscaling}a.), mixing in laser light with carefully selected intensity provides a higher SKR.

Despite the moderate brightness of our QDS implementation, the high purity enables long distance communication for BB84 with a SKR of $5 \times 10^{-6}$ bits per pulse at approximately 30 dB of channel losses (equivalent to few hundreds bits per second over 140 km of standard telecom fiber with linear attenuation $\alpha$ = 0.21 dB/km). At low distances, where the impact of count rate is more consequential, the SKR can be improved by mixing in laser light. In this configuration, adding laser increases the rate of events detected at the receiver and surpasses the higher cost of privacy amplification even though the single-photon purity is degraded. This leads to an overall higher secret key rate, asymptotically converging to the scaling that would be observed using laser pulses only, at equivalent distances.

To theoretically understand the measured scaling data, we developed a proper statistical description of the incoherent mixture of the QD-generated single photons and the Poissonian statistics of the laser is necessary. The single-photon statistics from the quantum dot source can be described by a mixed Fock state $\hat{\rho}_{\text{QD}} = p_0|0\rangle\langle0| + p_1|1\rangle\langle1| + p_2|2\rangle\langle2|$ involving only three non-zero components $p_0$, $p_1$, and $p_2$, with $p_2 \ll p_1$, where the coherence between photon-number components has been erased by its incoherent phonon-assisted off-resonant excitation. This photon-number mixture also prevents any eavesdropper from performing a coherent attack on this part of the photon stream. The exact value of the full photon-number distribution can be determined with two parameters: the single-photon purity (calculated using the $g^{(2)}(0)$) and the collected brightness. For realistic deployments, a phase scrambling procedure with the laser pulses should be implemented to erase photon-number coherence for every photon of the mixed statistics \cite{lo_phase_2005} for instance using a phase modulator and discrete phase randomization \cite{cao_discrete-phase-randomized_2015, tang_experimental_2014}. We thus assume that the Poissonian statistics of the laser with mean photon number $\mu_{\text{laser}}$ corresponds to the incoherent Fock state mixture:
\begin{equation}
    \hat{\rho}_{\text{laser}} = e^{-\mu_{\text{laser}}} \sum_{n = 0}^{+\infty} \frac{\mu_{\text{laser}}^n}{n!} |n\rangle\langle n|\,.
\end{equation}

The photon-number distribution of the mixture of laser and single photons is defined as the joint probability distribution of both statistics. For instance, the probability for Alice to send two photons $p_{\text{2, mixed}}$ is given by three contributions: emitting either (i) two photons from laser light or (ii) from QD, and (iii) emitting a single photon from each of the sources. Hence:
\begin{equation}
    p_2^{\text{hybrid}} = e^{-\mu_{\text{laser}}} \left( p_0 \frac{\mu_{\text{laser}}^2}{2} + p_1\mu_{\text{laser}} + p_2 \right)\,.
\end{equation}

More details on the theoretical model can be found in the Supplementary Material.
Using this statistical modeling with the parameters extracted from independent experiments enables to predict the optimal SKR distance scaling including the single-photon--laser mixture, as shown in Fig. \ref{fig:optimizedscaling}a. In Fig. \ref{fig:optimizedscaling}b, we observe perfect agreement between the measured and simulated single-photon purity and optimal mixing parameter. The model proposed above thus accurately allows predicting directly the mixing which would optimize the SKR at a given distance for practical implementations.

\begin{figure*}
    \centering
    \includegraphics[scale = 0.50]{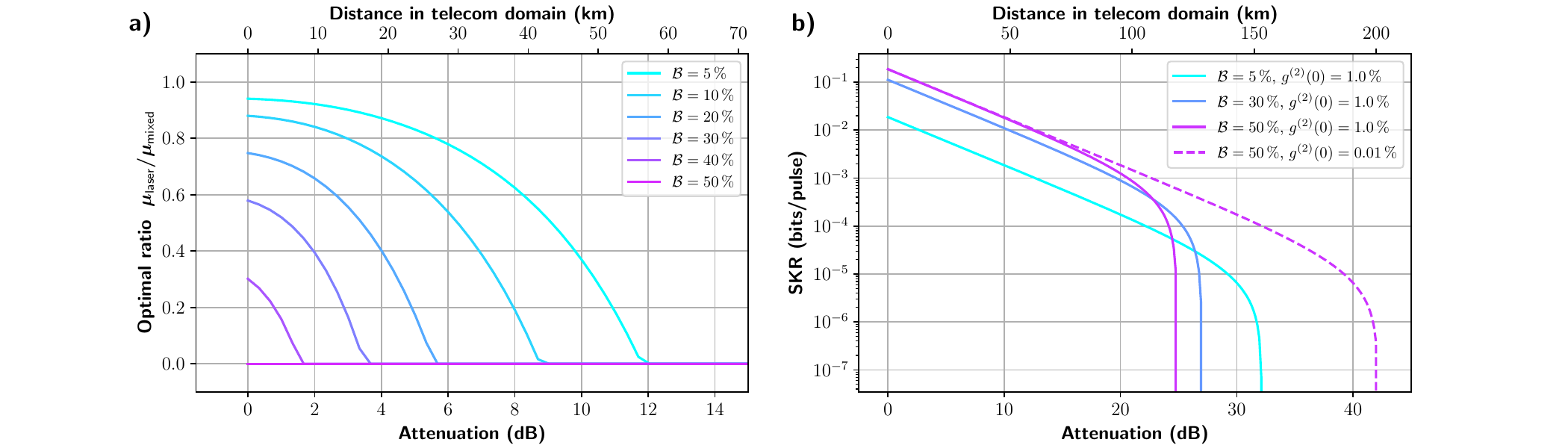}
    \caption{\textbf{Optimal mean photon number ratio for different collected brightnesses and corresponding distance scaling scenarios. a) } Simulated optimal relative mean photon ratio of the laser leading to the best SKR at various distances. \textbf{b) } Comparison of the simulated SKR distance scaling obtained with single photons for different configurations. Ths SKR obtained with a QDS with $g^{(2)}(0) = 1\,\%$ with different collected brightnesses is plotted as a function of attenuation (solid lines), as well as the SKR obtained with a QDS in the case where $g^{(2)}(0) = 0.01\,\%$ with collected brightness of 50\,\%.}
    \label{fig:advantage}
\end{figure*}

\subsection*{Towards an unconditional advantage of QDS}

We have demonstrated the possibility to increase the SKR in short-distance BB84 owing to an optimal mixing of laser light with single photons. This is performed in the regime where the single photon collection is still limited. Yet, our statistical modeling, as described above, allows us to identify a clear scenario of conditional and unconditional advantages induced by the laser mixing in scenarios involving SPS with higher collected brightnesses.

In Fig. \ref{fig:advantage}a, we vary the collected brightness: we observe the brighter the QDS, the less need there is for additional photons coming from the laser to maximize the SKR at a given distance. Importantly, the attenuation threshold also shifts gradually towards zero when increasing this efficiency. Assuming the same single-photon purity of the QDS measured in our experiment ($g^{(2)}(0) = 1.2\,\%$) an unconditional advantage of the single-photon statistics is obtained for collected brightnesses above 45.57\,\%, as the mixing with laser no longer improves the SKR at any attenuation nor does it improve the achievable distance for which a secret key can be distributed. This is achieved numerically in our simulation by optimizing the proportion of laser in the hybrid statistics. It is to be noted that this threshold is different from the collected brightness above which QDS reach higher SKR than laser at low distance. Thus, there is a range of collected brightnesses for which the best SKR at every distance is reached by the hybrid statistics. For instance, an ideal QDS ($g^{(2)}(0) = 0$) with ideal detection ($e_\text{d} = 0$) reaches higher SKR than laser for collected brightnesses above $e^{-1} \approx 36.79 \,\%$, yet the hybrid statistics still outperform both if this efficiency is below 50\,\%.

Until now, we considered the concept of advantage only regarding the SKR of each statistics and compared the cases of different collected brightnesses. Another important figure of merit is the maximum distance at which a secret key can be distributed. The impact of single-photon purity and collected brightness is investigated in the following in the case of single-photon statistics, as it is enabling positive SKR for long distances. In Fig. \ref{fig:advantage}b, we compare the simulated SKR distance scaling using a QDS only, for various levels of single-photon purity. Here, we observe that an increase in collected brightness without a corresponding improvement in single-photon purity is actually detrimental in terms of distance scaling because of the increased likelihood of sending two-photon Fock states into the quantum channel. Differently from what is done in \cite{bozzio_enhancing_2022}, we interpret the QDS and the collection setup as a black box device with a certain collected brightness and single-photon purity and we do not propagate the collection efficiency from the source to the setup. This approach helps to have a direct connection with the measured physical quantities that are not just accounted as an expression of losses. Indeed, if done rigorously with the definition of brightness and single-photon purity it can be shown that losses (accounted with a transmissivity $\eta$) are not just expressed as a scaling factor of the brightness but they can be inferred with the following formula:
\begin{equation}
    \mathcal{B} = \eta \mathcal{B}_0 + \eta (1 - \eta) \frac{1 - \mathcal{B}_0 g^{(2)}(0) - \sqrt{1 - 2\mathcal{B}_0 g^{(2)}(0)}}{g^{(2)}(0)} \neq \eta \mathcal{B}_0\,,
\end{equation}
with $\mathcal{B}_0$ being the brightness before losses. We emphasize here the paramount role played by the single-photon purity jointly with the collected brightness in order to ensure reliable QKD over long distances with QDS.

\section*{Discussion}

In this article, we have built a laboratory demonstrator to benchmark the BB84 protocol using an incoherently driven quantum dot–cavity single-photon source, comparing its SKR and distance performance with a weak Poissonian source. At short distances, more SKR is transmitted using the weak laser pulses. However, at longer distances, the single photons transmitted more SKR with up to approximately 30 dB distance. To leverage the advantage of both fields, we propose an adaptive method based on mixing the optimal amounts of laser light to the single-photon stream to optimize the SKR at short distances. This method, relying only on mixing the correct amount of laser light into the single-photon stream,  can be flexibly implemented with any single-photon source and, in principle, can be further combined with more advanced techniques exploring for instance optimization of the excitation scheme \cite{vyvlecka_robust_2023} or post-selection of the detection window \cite{kupko_tools_2020} to additionally extend the secured communication distance.

In practice, in the specific case of off-resonant optical excitation, hybrid statistics can be obtained in a simpler experimental implementation by tuning the spectral filtering of the excitation laser to mix in the laser directly instead of using an additional beam-splitter. A similar method could be implemented in the case of resonant excitation schemes. In this case, the excitation laser is rejected via a cross-polarization set-up, with an extinction ratio that could be tuned to generate hybrid statistics directly in its collection path. Achieving a real on-field implementation using this experimental technique still requires additional steps on the transmitter side. A fast active modulation of the polarization is necessary to prepare the four BB84 states in a way that matches the generation rate of the photons. This could be implemented with an electro-optic modulator with a high enough bandwidth. Benefiting from the low attenuation of standard telecom fibers is possible using quantum dots directly emitting in the telecom C-band \cite{nawrath_bright_2023, holewa_high-throughput_2024}. Alternatively, a quantum frequency conversion module from 925 nm to 1550 nm using a PPLN crystal could be used \cite{morrison_bright_2021, da_lio_pure_2022}.

\section*{Data availability}

Data is available from the authors upon reasonable request.

\putbib

\end{bibunit}

\section*{Acknowledgments}

This work was conducted within the research program of the QDLight joint laboratory (C2N/Quandela). It was partially supported the European Union’s Horizon 2020 FET OPEN projects QLUSTER (Grant ID 862035) and PHOQUSING (grant ID 899544), Horizon-CL4 program under the grant agreement 101135288 for EPIQUE project, the French RENATECH network, by the European Commission as part of the EIC accelerator program under the grant agreement 190188855 for SEPOQC project, the Plan France 2030 through the projects ANR22-PETQ-0011 and ANR-22-PETQ-0013. T.H. is part of the Max Planck School of Photonics supported by the German Federal Ministry of Education and Research (BMBF), the Max Planck Society and the Fraunhofer Society. F.R. acknowledges support from NSF (NSF grant No. ERC-1941583).

\section*{Author contributions}

Y.P. built the BB84 proof-of-concept, performed the measurement and wrote the code to process the data, relying on previous work by J.R.A., T.H. and L.L. Pe.S. and D.A.F. assisted in the preparation of the experiment. A.L. and M.M. grew the QDS used in this article. The nano-processing of the sample was made thanks to N.S. D.A.F and P.S. are the principal investigators of this work conceived by D.A.F. with inputs from F.R. Y.P., Pe.S. and D.A.F. developed the theoretical model of the hybrid statistics and wrote the manuscript. J.R.A., T.H., L.L., F.R. and P.S. gave feedback on the manuscript.

\section*{Competing interests}

N.S. is co-founder and CEO of the company Quandela. P.S. is co-founder and scientific advisor of the same company. All other authors declare no competing interests.

\clearpage

\onecolumngrid

\section*{Supplementary Material}

\begin{bibunit}

\subsection{Experimental details}

\subsubsection{Quantum dot single-photon source}

The single photons are generated by a self-assembled InGaAs quantum dot produced by growing a layer of InAs on GaAs matrix with molecular beam epitaxy (MBE). Because of the lattice mismatch between the two materials inducing strain, randomly distributed InAs nanostructures appear in a process called Stranski-Krastanov growth. The planar cavity is then placed between two sets of distributed Bragg reflectors (DBR) and etched in a micropillar structure using \textit{in-situ} lithography \cite{dousse_scalable_2009}, producing a cavity with quality factor Q $\approx$ 10300. This allows us to benefit from the Purcell effect which significantly enhances the colleccted brightness of the photons. Additionally, the micropillar is electrically contacted, allowing to tune the QD in resonance with the cavity using the Stark effect \cite{nowak_deterministic_2014}.

In this experiment, single photons are produced using the neutral exciton transition of the QD used in this work. The excitation laser is linearly polarized to address exclusively one of the two excited states of the exciton, separated by a fine-structure splitting constant due to the asymmetry of the nanostructure.

\subsubsection{Estimation of the mean photon number}

The mean photon number $\mu_{\text{laser}}^{\text{Alice}}$ characterizing the Poissonian statistics at the transmitter point is estimated with the information of events measured by Bob and the transmission $\eta_0$ of the quantum channel from Alice to Bob without applying any voltage to the attenuator, accounting for losses in the fibers and connections. Bob measures events triggered by three different phenomena: (i) the emission of a photon from the QD source, (ii) from the laser, or (iii) a dark count triggering a detection without photon arrival. Since Bob uses threshold detectors, multiphoton states cannot be discriminated from single clicks. As we mix laser and QD light incoherently, and as dark counts are uncorrelated with Alice's encoded qubits, we assume all three phenomena to be mutually independent. The probability that Bob registers a click $p_{\text{click}}$ can hence be expressed as the inverse probability of not getting a click from any of these three independent events:
\begin{equation}
    p_{\text{click}} = 1 - \left( 1 - p_{\text{dc}} \right) \left( 1 - p_{\text{click}}^{\text{QD}} \right) \left( 1 - p_{\text{click}}^{\text{laser}} \right)\,,
\end{equation}
where $p_{\text{dc}}$, $p_{\text{click}}^{\text{QD}}$, and $p_{\text{click}}^{\text{laser}}$ are respectively the probabilities of recording an event triggered by dark counts, detection of a photon from the QD source, and detection of a photon from the laser source. When only looking at the fraction of photons emitted from Alice's laser, their statistics remains Poissonian when arriving to Bob, with mean photon number $\mu_{\text{laser}}^{\text{Bob}} = \eta_0 \mu_{\text{laser}}^{\text{Alice}}$, thus:
\begin{equation}
   1 - p_{\text{click}}^{\text{laser}} = e^{-\eta_0 \mu_{\text{laser}}^{\text{Alice}}}\,.
\end{equation}
The mean photon number of the laser at Alice's end can therefore be inferred by measuring the dark counts, the total events and the events triggered by QDS emissions, as:
\begin{equation}
    \mu_{\text{laser}}^{\text{Alice}} = -\frac{1}{\eta_0} \ln \left( \frac{1 - p_{\text{click}}}{\left( 1 - p_{\text{dc}} \right) \left( 1 - p_{\text{click}}^{\text{QD}} \right)} \right)\,.
\end{equation}

Furthermore, the estimation of the mean photon number coming from the QDS is estimated with the measurement of counts and single-photon purity at Alice's end. With those quantities, the total mean photon number $\mu_{\text{mixed}}^{\text{Alice}} = \mu_{\text{QD}}^{\text{Alice}} + \mu_{\text{laser}}^{\text{Alice}}$ can be computed, along with the ratio of mean photon number assessing the balance between laser and pure single photons in the states sent by Alice.

\subsubsection{Calibrating spectral filters}

\begin{figure}[ht]
    \centering
    \includegraphics[scale = 0.45]{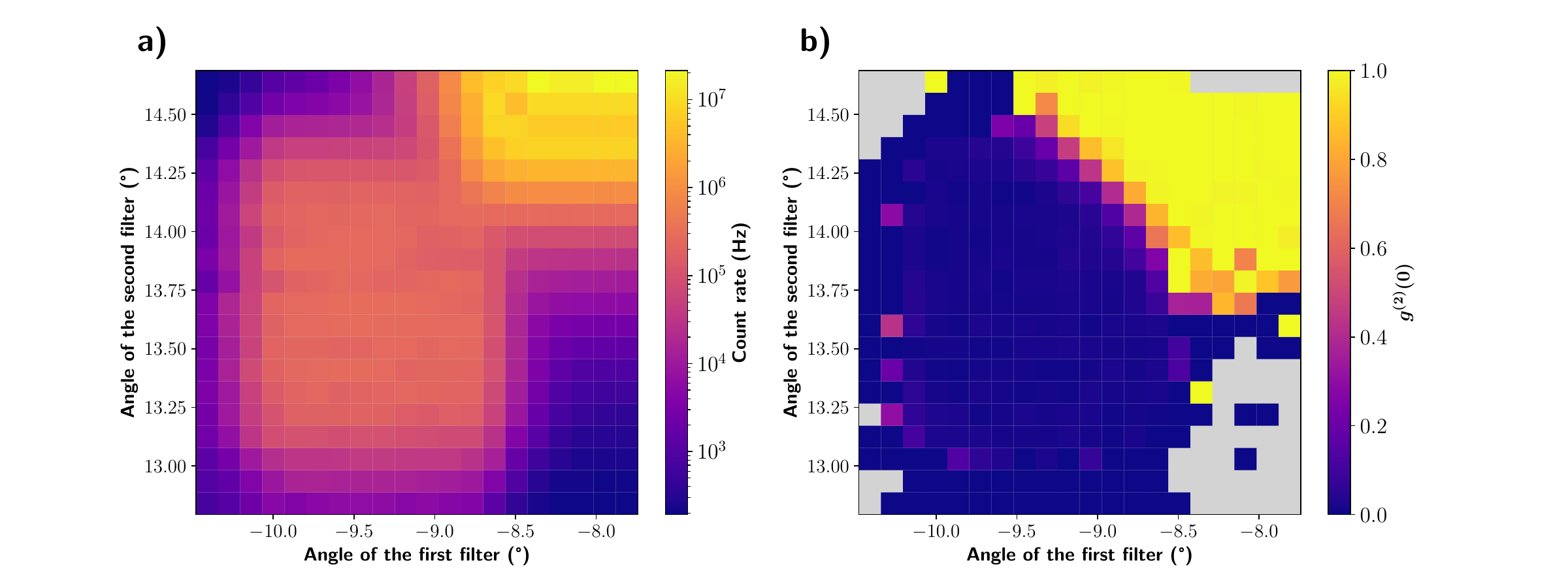}
    \caption{\textbf{Scanning parameters to compute the secret key rate (SKR).} Two parameters are measured for 20 $\times$ 20 different sets of filter positions (20 positions each): \textbf{a)} the raw count rate at Bob's end and \textbf{b)} the second-order autocorrelation function at time delay zero $g^{(2)}(0)$. The grey areas indicate positions for which the computation of the $g^{(2)}(0)$ was not possible either due to the saturation of the SNSPDs or to a low count rate.}
    \label{fig:colormaps}
\end{figure}

Out of the three spectral filters used to reject the excitation laser and collect only single photons, two are mounted on Thorlabs Elliptec ELL18K rotation stages to adjust their tilt angles in a programmable way. By tilting the filters, we are changing the transmission efficiencies of both the wavelength of the single photons and the excitation laser. This entails a capability to find the best position for the filters in terms of count rate and single-photon purity (Fig. \ref{fig:colormaps}) of the photon stream at the output, while also being able to mix in laser light and single photons directly in the same optical path.

\subsection{Derivation of theoretical bounds and models}

\subsubsection{Security analysis of practical BB84 protocol}

We remind here the theoretical work on practical BB84 leading to the secret key rate formula \cite{ma_quantum_2008, lo_decoy_2005, gottesman_security_2004}. We first define the yield $Y_k$ of a Fock state $|k\rangle$ as the conditional probability of having detected an event on Bob's detectors given that Alice sent a state containing $k$ photons. Explicitly:

\begin{equation}
    Y_k = Y_0 + \left(1 - Y_0 \right) \left[ 1 - \left(1 - \eta \right)^k \right]\,,
    \label{eq:yield}
\end{equation}

with $\eta$ being the general transmission factor between Alice and Bob and $Y_0$ the dark count probability. This equation can be understood as the sum of the two cases leading to Bob recording a detection: either Bob detects a dark count with probability $Y_0$, or at least one of the $k$ photons sent by Alice arrives to Bob with probability $ \left[ 1 - \left(1 - \eta \right)^k \right]$.

The source used for BB84 is characterized by its photon-number distribution ${p_k}$. We define the gain $Q_k$ of a state containing $k$ photons as:

\begin{equation}
    Q_k = p_k Y_k\,,
    \label{eq:gain}
\end{equation}

which is the probability for Alice to send a state containing $k$ photons that will be recorded by Bob. We also introduce the QBER generated by $k$-photon states as:

\begin{equation}
    e_k = \frac{e_0 Y_0 + e_{\text{d}} \left[1 - \left(1 - \eta \right)^k \right]}{Y_k}\,,
    \label{eq:error}
\end{equation}

where $e_{\text{d}}$ is the probability for each photon to enter the wrong detection path. Finally, we define the total gain and the total QBER as the sum of gains and QBER due to $k$-photon states, with:

\begin{equation}
    Q_\text{tot} = \sum_{k = 0}^{+\infty} Q_k
    \quad \text{and} \quad
    E_\text{tot} = \frac{1}{Q_{\text{tot}}}\sum_{k = 0}^{+\infty} e_k Q_k\,.
    \label{eq:total}
\end{equation}

By definition, the total gain $Q_{\text{tot}}$ is the total number of photons arriving on Bob's detectors, and is refered in the main text as $p_{\text{click}}$. From there, the SKR can be bounded by:

\begin{equation}
    \text{SKR} \geqslant \frac{1}{2} \left( Q_{k < 2} \left(1 - H_2(e_{k < 2}) \right) - f_\text{EC}Q_\text{tot}H_2(E_\text{tot}) \right)\,,
    \label{eq:SKRsup}
\end{equation}

with $f_{EC}$ the error correcting factor and $H_2(x) = - x \log_2(x) - (1 - x) \log_2(1 - x)$ the binary entropy function, $Q_{k < 2} = Q_0 + Q_1$ and $Q_{k < 2} e_{k < 2} = Q_0 e_0 + Q_1 e_1$. However, this bound cannot be used as it is in practise. On the experimental side, one only has access to $Q_\text{tot}$ and $E_\text{tot}$, not $Q_1$ nor $e_1$ nor $Q_0$ nor $e_0$, therefore needing some estimation of both parameters. In the pessimistic assumption that Eve has control over the entire quantum channel, they can choose to block every single-photon state and let every multi-photon state pass through, as it is the best configuration to perform their attack. This assumption draws back to consider that $Y_k = 1$ and $e_k = 0$ for all $k \geqslant 2$. In this scenario, one can make the following estimation of $Q_{k < 2}$ and $e_{k < 2}$ using Eq. \ref{eq:total}:
\begin{equation}
    Q_{k < 2} = Q_\text{tot} - \sum_{k = 2}^{+\infty} p_k
    \quad \text{and} \quad
    e_{k < 2} = \frac{E_\text{tot} Q_\text{tot}}{Q_{k < 2}}\,.
\end{equation}
It can be now noticed that $Q_\text{tot} - \sum_{k = 2}^{+\infty} p_k = p_{\text{click}} - p_{\text{m}} = p_{\text{click}}A$ with the single-photon component $A$ defined in the main text, therefore yielding Eq. \ref{eq:SKR}.

\subsubsection{Poisson-distributed sources}

Depending on the statistics of the source that is used by Alice as an emitter for the BB84 protocol, the photon-number distribution will change. Lasers behave as Poisson-distributed sources, their statistics is therefore defined by a single parameter $\mu$, the mean photon number. In this case:
\begin{equation*}
    p_k = e^{-\mu}\frac{\mu^k}{k!}\,.
\end{equation*}

\subsubsection{Single-photon sources}

In the case of our single-photon source, the single-photon purity is very high with $\mathcal{P} = 1 - g^{(2)}(0) = 98.8\,\%$, hence we approximate the photon-number distribution of the quantum-dot-emitted photons by truncating it to the states containing no more than two photons. Additionally, since we are taking advantage of the off-resonant, phonon-assisted excitation scheme to generate the photons without having coherence in the photon-number basis, an incoherent mixture is produced that is described by the mixed Fock state:
\begin{equation}
    \hat{\rho}_{\text{QD}} = p_0|0\rangle\langle0| + p_1|1\rangle\langle1| + p_2|2\rangle\langle 2|\,.
\end{equation}

The components of the photon-number distribution $p_0$, $p_1$, and $p_2$ can be determined at one point of the photons' path with the measurement of both the number of events on a threshold single-photon detector $\mathcal{B} = p_1 + p_2$, the $g^{(2)}(0)$, and the normalization condition $p_0 + p_1 + p_2 = 1$. In general:
\begin{equation}
   \mu^2 g^{(2)}(0) = \sum_{n \geqslant 2} n (n-1) p_n\,,
\end{equation}
with $\mu = \sum_{n \geqslant 1} n p_n$ the mean photon number. In the case of the truncated photon-number distribution, this writes as:
\begin{align*}
   \left(p_1 + 2 p_2 \right)^2 g^{(2)}(0) &= 2 p_2 \\
   \left(\mathcal{B} + p_2 \right)^2 g^{(2)}(0) &= 2 p_2\,.
\end{align*}
Solving the quadratic equation for $p_2$ and discarding the solution incompatible with a single-photon source ($g^{(2)}(0) > 0.5$), the probability of generating a 2-photon state is:
\begin{equation}
    p_2 = \frac{1 - g^{(2)}(0)\mathcal{B} - \sqrt{1 - 2g^{(2)}(0)\mathcal{B}}}{g^{(2)}(0)}\,.
\end{equation}

\subsubsection{Hybrid statistics}

By using both pure single photons coming from the QD source and laser pulses, in which photons follow a Poissonian statistics, we are creating a hybrid photon-number distribution. The two kinds of photons are mixed incoherently, making the hybrid distribution the joint probability distribution from the pure photons statistics and the laser statistics. To measure vacuum from this distribution means that no photon was produced either by the QD source or the laser source, which gives $p_0^{\text{hybrid}} = p_0 e^{-\mu_{\text{laser}}}$. Similarly, to produce one photon from the hybrid distribution means producing one photon from the QD or the laser exclusively, yielding $p_1^{\text{hybrid}} = p_1 e^{-\mu_{\text{laser}}} + p_0 \mu_{\text{laser}} e^{-\mu_{\text{laser}}}$. The probability of producing an $n$-photon state for $n \geqslant 2$ can be written in a general way as:
\begin{equation}
    p_n^{\text{hybrid}} = e^{-\mu_{\text{laser}}} \left( p_0 \frac{\mu_{\text{laser}}^n}{n!} + p_1 \frac{\mu_{\text{laser}}^{n-1}}{(n-1)!} + p_2 \frac{\mu_{\text{laser}}^{n-2}}{(n-2)!} \right)\,.
\end{equation}

The mean photon number of this hybrid statistics is:
\begin{align*}
    \mu_{\text{mixed}} &= \sum_{n \geqslant 1} n p_n^{\text{hybrid}} \\
    &= p_1^{\text{hybrid}} + e^{-\mu_{\text{laser}}} \sum_{n \geqslant 2} n \left(p_0 \frac{\mu_{\text{laser}}^n}{n!} + p_1 \frac{\mu_{\text{laser}}^{n-1}}{(n-1)!} + p_2 \frac{\mu_{\text{laser}}^{n-2}}{(n-2)!} \right) \\
    &= p_1 e^{-\mu_{\text{laser}}} + p_0 \mu_{\text{laser}} e^{-\mu_{\text{laser}}} + e^{-\mu_{\text{laser}}} \left( p_0 \mu_{\text{laser}} \left( e^{\mu_{\text{laser}}} - 1 \right) + p_1 \left( \mu_{\text{laser}} e^{\mu_{\text{laser}}} + e^{\mu_{\text{laser}}} - 1 \right) + p_2 e^{\mu_{\text{laser}}} \left( \mu_{\text{laser}} + 2 \right) \right) \\
    &= p_1 e^{-\mu_{\text{laser}}} + p_0 \mu_{\text{laser}} e^{-\mu_{\text{laser}}} + p_0 \mu_{\text{laser}} \left( 1 - e^{-\mu_{\text{laser}}} \right) + p_1 \left( \mu_{\text{laser}} + 1 + e^{-\mu_{\text{laser}}} \right) + p_2 \left( \mu_{\text{laser}} + 2 \right) \\
    &= p_0 \mu_{\text{laser}} + p_1 \left( \mu_{\text{laser}} + 1 \right) + p_2 \left( \mu_{\text{laser}} + 2 \right) \\
    &= \mu_{\text{laser}} \left( p_0 + p_1 + p_2 \right) + p_1 + 2 p_2 \\
    &= \mu_{\text{laser}} + \mu_{\text{QD}}\,.
\end{align*}

As the hybrid statistics is the joint probability distribution of the QD source distribution and the Poissonian distribution, its mean photon number is the sum of both mean photon numbers.

\subsubsection{Advantage threshold for hybrid statistics in the ideal case}

We derive here the following claim from the main text: in the case of perfect single-photon purity, single-photon statistics perform better than Poissonian statistics for a collected brightness above 36.79\,\%, and hybrid statistics outperform both when this brightness is under 50 \,\%, above which the mixing strategy does not bring an advantage.

At 0 dB attenuation, when the error rate is equal to 0, the secret key rate draws down to being equal to the probability of generating single-photon states for each statistics. Hence, if $p_1$ is the single-photon probability for the QDS and $\mu_{\text{laser}}$ the mean photon number of the laser:
\begin{align}
    \text{SKR}_{\text{QD}} &= p_1 \\
    \text{SKR}_{\text{laser}} &= \mu_{\text{laser}} e^{-\mu_{\text{laser}}} \\
    \text{SKR}_{\text{hybrid}} &= p_1 e^{-\mu_{\text{laser}}} + (1 - p_1) \mu_{\text{laser}} e^{-\mu_{\text{laser}}}
\end{align}

Maximizing the SKR for laser under the condition that $\mu_{\text{laser}}$ is positive yields the threshold of $e^{-1} \approx 36.79\,\%$. The same reasoning applied to the SKR of the hybrid statistics shows that we can find an optimal $\mu_{\text{laser}}$ so that the hybrid statistics performs the best, as long as $p_1 \leqslant 0.5$.

\subsection{Exploration of practical scenarios}

\subsubsection{Hybrid statistics with higher error rates}\label{higher_errors}

\begin{figure}[ht]
    \centering
    \includegraphics[scale = 0.8]{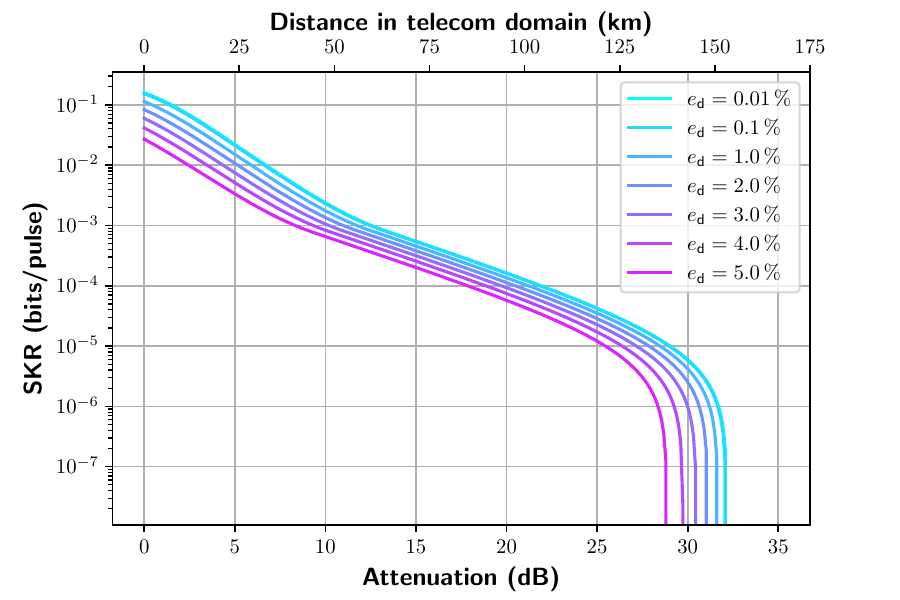}
    \caption{\textbf{Optimized scaling in the case of hybrid statistics for several error rates.} For each attenuation, the optimal secret key rate is computed over the mean photon number of the laser in the hybrid statistics. The QD source parameters are set to match the source used in the experiment, with $g^{(2)}(0)$ = 1.2\,\% and a collection efficiency at the end of Alice of 4.09\,\%. The SKR distance scaling is computed for different scenarios corresponding to several values of error rates from 0.01\,\% to 5 \,\%.}
    \label{fig:errors}
\end{figure}

In our experiment, we achieved a 0.8\,\% error rate by considering only the errors made on the basis $\left\{ |H\rangle, |V\rangle \right\}$, assuming that this rate would be similar in the other measurement basis $\left\{ |D\rangle, |A\rangle \right\}$. In order to investigate more practical implementations with a higher error rate, the optimized SKR obtained with hybrid statistics, in the case of different error rates, is shown in Fig. \ref{fig:errors}. Although a lower error rate means a better distance scaling and higher SKR unconditionally, the gain in maximum achievable distance from a 2\,\% to 0.01\,\% error rate is rather modest with an increase of a few dB which would correspond to approximately 10 km to 20 km in telecom fiber.

\subsubsection{Single-photon sources with higher brightnesses}

\begin{figure}[ht]
    \centering
    \includegraphics[scale = 0.8]{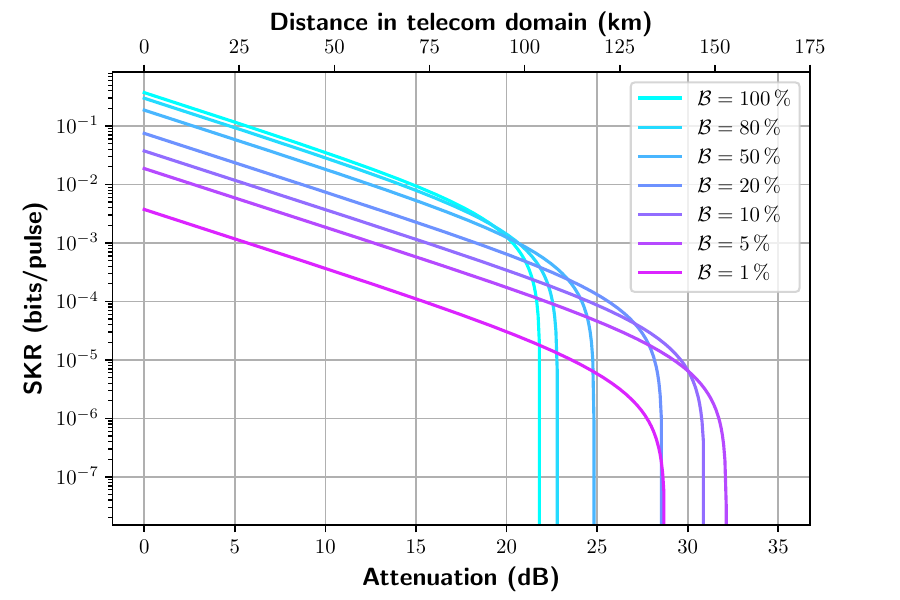}
    \caption{\textbf{SKR distance scaling for single-photon sources with different collected brightnesses.} Simulation curves are shown in the case of a single-photon source generating no states containing more than two photons, with $g^{(2)}(0)$ = 1\,\%. The SKR is plotted as a function of the attenuation for collected brightnesses going from 1\,\% to perfect collected brightness of 100\,\%.}
    \label{fig:efficiencies}
\end{figure}

The simulation results presented in Fig. \ref{fig:efficiencies} lead to a behaviour that might seem counter-intuitive at first. In the case of single-photon sources with a fixed, high single-photon purity (here $g^{(2)}(0) = 1\,\%$), one could imagine that we always have an interest in increasing the collected brightness of the source as much as possible. Yet, when looking at the distance scaling of the SKR in different scenarios, it appears clearly that the longest distance is not reached with the highest collected brightness. Even though high collected brightness increases the SKR for low distances, it makes the SKR decay faster with distance, as it makes Alice send more states containing more than one photon. This simulation highlights inadequacy of increasing the collected brightness of single-photon sources alone. The more photons are being produced and sent by Alice, the more stringent the requirement on the single-photon purity becomes. This effect, however, does not imply that a high single-photon purity enables reaching long distances regardless of its collected brightness, as it can be seen in Fig. \ref{fig:efficiencies}. Indeed, not being able to collect and hence prepare enough photons leads to a faster decay of the SKR.

\begin{figure}[ht]
    \centering
    \includegraphics[scale = 0.8]{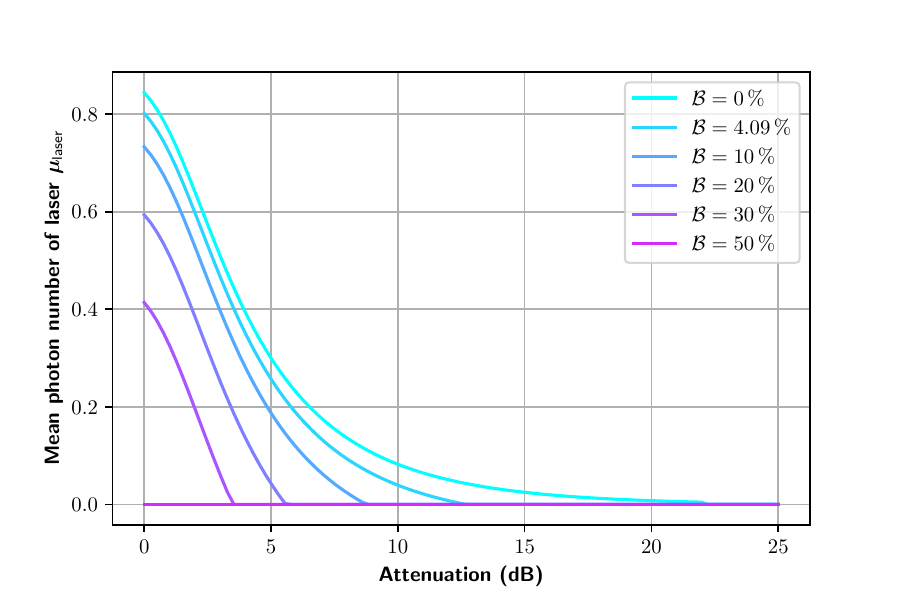}
    \caption{\textbf{Optimal mean photon of the laser for different collected brightnesses.} The mean photon number allowing to reach the highest SKR for each attenuation is plotted as a function of said attenuation in the scenario of different collected brightnesses for the QDS. The $0\,\%$ collected brightness corresponds to the case where only laser light is used to distribute a key.}
    \label{fig:OptiMu}
\end{figure}

\subsubsection{Optimal mean photon numbers for hybrid statistics}

We show in Fig. \ref{fig:OptiMu} that the amount of laser needed to reach optimal SKR with hybrid statistics is decreasing with the brightness of the QDS. It can be noticed that, in the case of our experiment with 4.09\,\% collected brightness, the optimal $\mu_{\text{laser}}$ is similar over the whole attenuation range to the case where only laser is used. This explains the nearly identical SKR in both cases for low attenuation. This simulation highlights again the potential of QDS with high collected brightnesses, as the need for laser is mitigated.

\putbib

\end{bibunit}

\end{document}